\documentclass[aps,prb,twocolumn,groupedaddress,showpacs]{revtex4}
\usepackage{graphicx}
\usepackage{amssymb}

\begin{document}

\preprint{}

\title{Illumination--induced changes of the Fermi surface topology 
in three--dimensional superlattices}
\author{N.\ A.\ Goncharuk}
\affiliation{Institute of Physics, Academy of Science of the Czech
Republic,\nolinebreak[5] v.v.i.,\\ 
Cukrovarnick\'{a} 10, 162 53 Praha  6, Czech Republic}
\author{L.\ Smr\v{c}ka}
\affiliation{Institute of Physics, Academy of Science of the Czech
Republic,\nolinebreak[5] v.v.i.,\\ 
Cukrovarnick\'{a} 10, 162 53 Praha  6, Czech Republic}
\author{P.\ Svoboda}
\affiliation{Institute of Physics, Academy of Science of the Czech
Republic,\nolinebreak[5] v.v.i.,\\ 
Cukrovarnick\'{a} 10, 162 53 Praha  6, Czech Republic}
\author{P.\ Va\v{s}ek}
\affiliation{Institute of Physics, Academy of Science of the Czech
Republic,\nolinebreak[5] v.v.i.,\\ 
Cukrovarnick\'{a} 10, 162 53 Praha  6, Czech Republic}
\author{J.\ Ku\v{c}era}
\affiliation{Institute of Physics, Academy of Science of the Czech
Republic,\nolinebreak[5] v.v.i.,\\ 
Cukrovarnick\'{a} 10, 162 53 Praha  6, Czech Republic}
\author{Yu.\ Krupko}
\affiliation{Grenoble High Magnetic Field Laboratory
CNRS, 25 Avenue des Martyrs  B.P.166, 38042 Grenoble cedex 9, France}
\author{W.\ Wegscheider}
\affiliation{University of Regensburg, 
Universit\"{a}strasse 31, D-93040 Regensburg, Germany}




\date{\today}

\begin{abstract}
The magnetoresistance of the MBE-grown GaAs/Al$_{0.3}$Ga$_{0.7}$As
superlattice with Si-doped barriers has been measured in tilted
magnetic fields in the as-grown state, and after brief illumination
by a red-light diode at low temperature, $T\approx 0.3$ K. A
remarkable illumination-induced modification of magnetoresistance 
curves has been observed, which indicates a significant change of 
the superlattice Fermi surface topology. Analysis of magnetoresistance 
data in terms of the tight-binding model reveals that not only 
electron concentration and mobility have been increased 
by illumination, but also the coupling among 2D electron layers 
in neighboring quantum wells has been reduced.
\end{abstract}

\pacs{73.43.Qt; 73.43.Jn; 02.60.Ed; 03.65.Sq}

\maketitle


\section{Introduction\label{Intro}}
Semiconductor superlattices (SLs), first proposed by Esaki and Tsu in
1970 \cite{Esaki}, are layered structures composed of regularly spaced
quantum wells (QWs) separated by barriers. A review of early research
of SLs was given, e.g., by Maan~\cite{Maan} and Helm.~\cite{Helm}

When doped by donors, the barriers provide electrons for
two-dimensional (2D) electron sheets inside the QWs. As discussed, e.g.,
in Refs.~\cite{Chang,Nachtwei} the resulting electronic structure of
the SL depends on the strength of the periodic potential as well as on
the level of the doping, and can range from a quasi-three-dimensional
(3D) system to a nearly 2D system of independent layers of 2D electron
gas in each QW.

For strong potential, usually provided by thick and high barriers, the
tunnel-coupling of electron layers, associated with overlap of wave
functions from neighboring wells, is negligible and SL minibands 
reduces to almost discrete energy levels. Moreover, in the
narrow wells, the minibands are well separated and usually only one
miniband is occupied. Then the electron system can be considered as
2D, and the Fermi surface (FS) is cylindrical in shape.

\begin{figure}[htb]
\includegraphics[width=0.85\linewidth,angle=0]{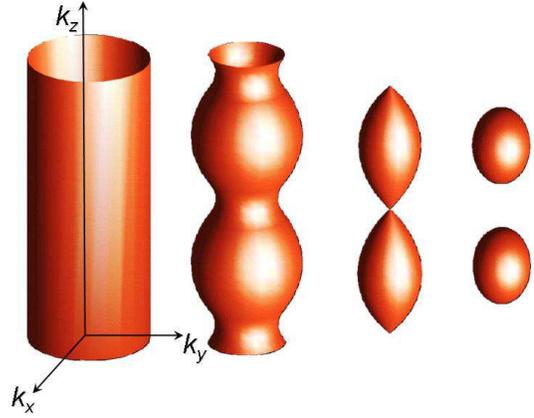}
\caption{\label{FS_topology}(Color online) The scheme of a variety of SL Fermi
surfaces. From the left to the right: a cylinder is related to $E_F$
lying well above the narrow SL miniband top (strictly 2D behavior); 
a corrugated cylinder corresponds to $E_F$ lying above the SL miniband top
(2D behavior); 
touching FSs arise when $E_F$ lies at the top of the
SL miniband and define the point of the 2D~$\rightarrow$~3D transition;
separated ovals occur when $E_F$ lies below the SL miniband top 
(strictly 3D behavior).}
\end{figure}

With decreasing the barrier thickness tunneling of electrons becomes
increasingly important, manifesting itself in the broadening of
minibands. The FS becomes the corrugated
cylinder if the Fermi energy $E_F$ is above the miniband top, or
closed semielliptical Fermi ovals (each fully contained within the
first Brillouin zone) for the Fermi energy below  the miniband
top. The latter electronic structure corresponds to the anisotropic
3D electron gas. The development of the FS topology with increasing 
tunnel-coupling between QWs is schematically shown 
in Fig.~\ref{FS_topology}. We call the change of the FS from an open 
corrugated cylinder towards closed ovals the 2D~$\rightarrow$~3D transition.

The SL Fermi surface topology can be determined experimentally by
magnetotransport measurements in magnetic fields $\vec{B}$ tilted with
respect to the QW planes.~\cite{Chang, Nachtwei, Stormer} The
quasiclassical theory predicts that an electron is driven by the
Lorentz force around orbits, defined in $k$-space by intersections of
the FS and planes perpendicular to the direction of the applied
magnetic field. In the case of closed orbits, the electron motion is quantized
into Landau levels, that manifest themselves through Shubnikov-de Haas
(SdH) oscillations observable in the longitudinal magnetoresistance.

The Onsager-Lifshitz relation \cite{Onsager} states that SdH
oscillations are periodic in $1/B$ with the period determined by the
extremal cross-sections of the FS.~\cite{Ashcroft}  By tilting the
sample, a number of cross-sections can be examined and the FS 
reconstructed. This is true only for closed FSs. For 
corrugated cylinders, open orbits appear for field orientations
close to the QW planes, and the SdH oscillations are suppressed.

An interesting case are SLs with the Fermi energy just below the top
of the miniband. The Fermi ovals nearly touch the Brillouin zone
boundaries and, therefore, the extremal closed orbits corresponding to
the in-plane magnetic fields also nearly touch each other.  If the
magnetic field is strong enough, the Onsager-Lifshitz quantization
scheme can be violated by the magnetic breakdown, and the closed
orbits effectively interconnected to the open ones. In that case we
can speak about {\em magnetic-field}-induced 3D$\rightarrow$2D transition.
At this point the quasiclassical description fails and the quantum-mechanical
treatment becomes necessary.

The deviations from the standard $1/B$ periodicity due to the magnetic
breakdown have been already reported in Ref.~\onlinecite{Chang}, for SdH
oscillations corresponding to the Landau levels with low indices,
measured in strong tilted magnetic fields. The magnetic-field-induced
3D$\rightarrow$2D transition was intentionally studied by Jaschinski
{\it et al.}~\cite{Jaschinski} Our experimental investigation of
this effect is described in Ref.~\onlinecite{PhysE_sl} and interpreted by the
full quantum-mechanical theory developed in Ref.~\onlinecite{PRB_sl}.

In the present publication we concentrate on another possibility: the
illumination-induced changes of the superlattice FS topology. 
The illumination of doped semiconductor structures at low temperatures 
leads to the persistent photoconductivity effect (PPC) and increases the
concentration of free carriers. As the increase of the concentration
means the increase of the Fermi energy, one expects that it might be
possible to change the FS topology from closed ovals to a corrugated
cylinder. This paper reports on the experimental study demonstrating
that the illumination of the SL sample 
can indeed change the electronic
structure of the SL and induce a transition from a 3D to 2D state. To
exclude the magnetic breakdown, 
we limit our experiments to lower magnetic fields, for which the
quasiclassical interpretation of magnetotransport measurements is
valid.


\section{Experiment \label{Exp}}
The sample employed in our experiments is the MBE-grown
SL (30 periods), which consists of 29 GaAs QWs of width $d_w
= 4.5$~nm, separated by $4$~nm wide Al$_{0.3}$Ga$_{0.7}$As
barriers. This gives the period $d_z$ of the SL equal to $ 8.5$~nm.
Each barrier is composed of the central Si-doped part $2.7$~nm thick,
which is flanked by $0.65$~nm thick undoped spacer layers on both
sides. The samples from the same wafer were used in our previous study
of the magnetic-field-induced 3D$\rightarrow$2D transition
\cite{PhysE_sl} in magnetic fields up to $26$~T.

The Hall bar samples have been patterned by means of the optical
lithography, equipped with evaporated AuGeNi contact pads and adjusted
to ceramic chip carriers. The width of their conducting channel was $w
= 400$~$\mu$m, the length of the sample was $L =
1000$~$\mu$m. The chip carrier with the sample has been attached to a
rotation plate in a $^3$He cryostat, that fitted the bore of a
superconducting magnet providing magnetic fields from 0 to $13$~T, the
range of fields for which the quasiclassical interpretation of SdH
oscillation is reasonable.~\cite{PhysE_sl}

The plate made it possible to rotate the sample to any angle between
the perpendicular $\left(\varphi = 0^{\circ}\right)$ and in-plane
$\left(\varphi = 90^{\circ}\right)$ field orientations.  Standard
lock-in technique at $f = 13$~Hz has been employed to measure
simultaneously both the longitudinal $\rho_{xx}(B)$ and Hall
$\rho_{xy}(B)$ resistances during sweeping the magnetic field up or
down. All the data presented below were taken at the temperature of
about $0.3$~K.

A red-light LED was installed next to the sample holder to facilitate
an ``in situ'' illumination of the cooled-down sample in order to
enhance the electron concentration and induce the PPC. Short
current pulses ($\approx 100$~msec) were subsequently applied to
the LED and the decrease of the zero-field resistance of the sample
due to the illumination was monitored. A few such pulses were
sufficient to get the resistance saturation. All the curves reported
below for the ``illuminated'' sample have been measured in this
saturated state.
\section{Theory: model calculations \label{Theory}}
\subsection{Tight-binding superlattice miniband}
It follows from the general consideration presented in the introduction
that the energy spectrum of the SL can be written as  
\begin{equation}
\label{3D_dispersion}
E\left(\vec{k}\right)=\frac{\hbar^2k_x^2}{2m^{\ast}}+
\frac{\hbar^2k_y^2}{2m^{\ast}}+E\left(k_z\right),
\end{equation}
where $m^{\ast}=0.067\,m$ denotes the effective mass of electrons
moving freely in the $(x,y)$ plane of the QWs, and
$E\left(k_z\right)$ is the dispersion relation of the miniband which
describes tunneling of electrons through the
barriers. $E\left(k_z\right)$ is a periodic function,
$E\left(k_z\right) =E\left(k_z+2\pi/d_z\right)$.
Together with the knowledge of $E_F$, Eq.~(\ref{3D_dispersion})
determines the shape of the SL Fermi surface.

The aim of this publication is to determine the FS of the SL
from measured periods of SdH oscillations in tilted magnetic fields.
Therefore we employ the simplest possible model of
$E\left(k_z\right)$ to minimize the number of fitting parameters
describing the miniband structure.

The natural choice is the tight-binding model in which
\begin{equation}
\label{miniband}
E\left(k_z\right)=-2t\cos{\left(k_zd_z\right)}\,,
\end{equation}
where $t$ is the coupling constant, associated with the matrix element
of the SL potential between the overlapping wave functions from
neighboring wells, which determines the miniband width.  The SL
growth parameters such as the effective barrier thickness, the barrier
height and the QW width are thus represented by this single constant
\cite{Hu}, and the shape of the FS is described by three parameters $t$,
$d_z$ and $E_F$.  It will be shown in subsequent sections that this
approximation of $E\left(k_z\right)$ yields a reasonable agreement
with experimental findings.

The cosine approximation for a single SL miniband gives the following
expressions for the electron concentration per layer:
\begin{equation}
\label{N_closed}
N = \frac{m^{\ast}}{\pi^2\hbar^2}
\left[E_F\arccos{\left(-\frac{E_F}{2t}\right)+\sqrt{4t^2-E_F^2}}\,\right],
\end{equation}
for a closed FS and
\begin{equation}
\label{N_open}
N = \frac{m^{\ast}}{\pi\hbar^2}E_F ,
\end{equation}
for an open FS. These expressions include the spin degeneracy.
According to Eqs.~(\ref{N_closed})-(\ref{N_open}), 
the knowledge of the electron concentration, which can be determined 
from the Hall magnetoresistance data, allows the prediction of the Fermi 
energy value $E_F$.

\subsection{Quasiclassical approach, perpendicular and tilted 
magnetic fields}
After application of an external magnetic field $\vec{B}$, the energy
spectrum is converted to a set of Landau levels. The orbits
corresponding to extremal cross-section areas, $A_k$, of the FS are
most important. According to the Onsager-Lifshitz quantization rule
\cite{Onsager}, the expression
\begin{equation}
\label{O-Lq}
A_k = \frac{2\pi|e|B}{\hbar}\left(n+\frac{1}{2}\right), \,\,\
n=0,1,2,\cdots ,
\end{equation} 
determines a set of magnetic fields at which Landau levels cross
the Fermi energy.

We have chosen the geometry where a uniform magnetic field applied in
the $(y,z)$ plane
$\vec{B}=\left(0,B\sin{\varphi},B\cos{\varphi}\right)$ is tilted by an
angle $\varphi$ with respect to the growth axis $z$, and its
orientation varies between perpendicular
$\left(B=B_z,\,\varphi=0^{\circ}\right)$ and in-plane
$\left(B=B_y,\,\varphi=90^{\circ}\right)$ configurations.

The relation between the period of SdH oscillations in reciprocal
magnetic fields, $\Delta(1/B)$, and $A_k$ can be written as
\begin{equation}
\label{O-L}
A_k = \frac{2\pi|e|}{\hbar\Delta(1/B)\,.}
\end{equation}

With the energy spectrum given by Eqs.~(\ref{3D_dispersion}) and
(\ref{miniband}), the extremal cross-section area in tilted magnetic
fields takes the form
\begin{eqnarray}
\label{Ak_tilt}
\lefteqn{A_k^{tilted}(\varphi) = \frac{\sqrt{2m^{\ast}}}{\hbar}}\\ 
& & \times 2 \int{\sqrt{
E_F-\frac{\hbar^2k'^2\cos^2{\varphi}}{2m^{\ast}}+
2t\cos{\left(k'\sin{\varphi}\,d_z\right)}
}}\,dk'\,,\nonumber
\end{eqnarray}
where 
\begin{equation}
\label{k'}
k'=\frac{k_y}{\cos{\varphi}}-k_{z0}\,\tan{\varphi}=
k_{z0}\,\cot{\varphi}-\frac{k_z}{\sin{\varphi}}\,.
\end{equation}
In Eq.~(\ref{k'}) $k_{z0}$ is the point at which the plane
perpendicular to the direction of the magnetic field cuts the axis
$k_z\,$. Two types of extremal cross-sections exist: $A_k^B$ for $k_{z0}=0$
corresponds to a ``belly'' orbit and $A_k^N$ for $k_{z0}=\pm\pi/d_z$
to a ``neck'' orbit.  A wide range of tilt angles $\varphi$ makes it
possible to determine the FS of the sample.

In strictly perpendicular magnetic fields, the ``belly'' and ``neck''
extremal orbits become circular, with cross-section areas given by
simple expressions
\begin{eqnarray}
\label{Ak_B}
A_k^B=\frac{2m^{\ast}\pi}{\hbar^2}\left(E_F+2t\right)\,,\\
\label{Ak_N}
A_k^N=\frac{2m^{\ast}\pi}{\hbar^2}\left(E_F-2t\right)\,.
\end{eqnarray}
For the closed FSs, $|k_z|$ is always less than $\pi/d_z$, and only
``belly'' orbits with areas $A_k^B$ occur.

It follows from Eq.~(\ref{O-Lq}) that in the perpendicular magnetic
field the quantized energies corresponding to either the ``belly'' or
`neck'' orbit cut the Fermi energy at magnetic fields given by equations
\begin{eqnarray}
\label{B-EF}
\frac{\hbar|e|B^B}{m^{\ast}}\left(n+\frac{1}{2}\right)=E_F+2t,\\
\frac{\hbar|e|B^N}{m^{\ast}}\left(n+\frac{1}{2}\right)=E_F-2t.
\end{eqnarray}
Note that we use the energy scale with the origin in the middle of the
SL miniband, i.e., the miniband bottom lies at $E=-2t$ and its top at
$E=2t$.
\section{Experimental results \label{Results}}
\begin{figure}[!h]
\includegraphics[width=0.95\linewidth]{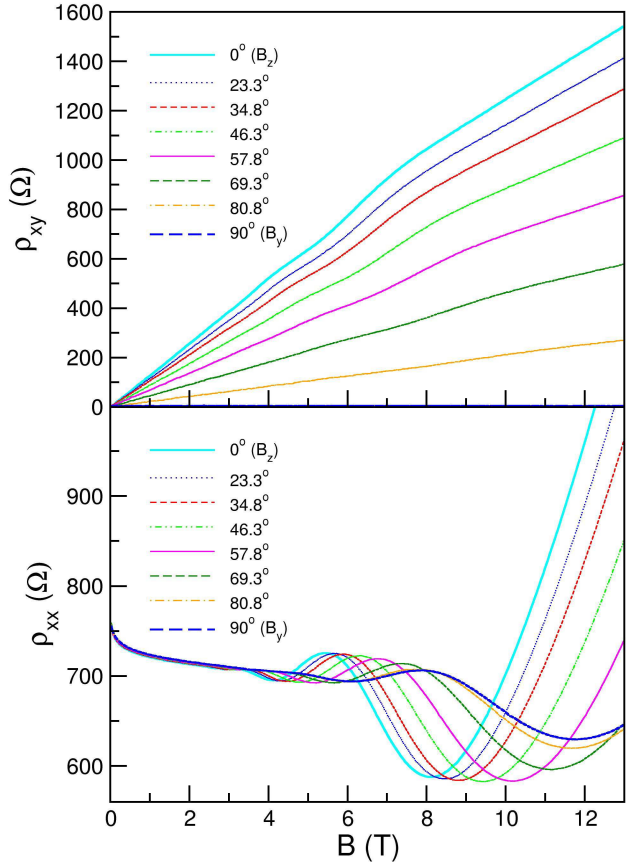}
\caption{\label{fig_2} 
(Color online) The experimental curves of the transversal 
$\rho_{xy}$ and longitudinal $\rho_{xx}$ magnetoresistance measured 
at $T = 0.3$~K for the set of tilt angles $\varphi$ of the applied 
magnetic field on the SL sample ``in dark''.} 
\end{figure}

Experimental data collected for the SL before illumination are
displayed in Fig.~\ref{fig_2}. Due to quasi-3D nature of its
electronic structure, quantum Hall plateaus are not distinguished on
Hall magnetoresistance curves $\rho_{xy}$.  The spin unresolved SdH
oscillations can be noticed on all $\rho_{xx}$ curves, including those
obtained in strictly in-plane magnetic fields. This confirms beyond
question that the FS is a closed semielliptic oval. The oscillations
have been found to be periodic in $1/B$ and a single period is well
detectable for all angles.

Knowing the number of QWs in the SL, we can determine the electron
concentration per 2D layer. It follows from the Hall magnetoresistance
measured in low perpendicular magnetic fields (under 3 T), that
$N_{d,H}\backsimeq 1.656\times 10^{11}$~cm$^{-2}$.

The experimental periods deduced from the SdH oscillations were put
into Eq.~(\ref{O-L}),  to calculate the angular dependence of
the extremal cross-section area $A_k^B(\varphi)$. The nonlinear fitting of
Eq.~(\ref{Ak_tilt}) to the experimentally determined curve gives the
parameters of the tight-binding model.  Excellent fit is possible with
values $t=4.6$~meV, $E_F=5.3$~meV.  This implies that the SL miniband
width is $4t=18.4$~meV and that the $E_F$ lies $3.9$~meV bellow the
top of the miniband.

The fitted values can be used in Eq.~(\ref{N_closed}) to calculate the
electron concentration from the period of the SdH oscillations. We get
$N_{d,SdH}\backsimeq 1.7\times 10^{11}$~cm$^{-2}$ in very good
agreement with $N_{d,H}$. The mobility calculated from the zero-field
resistance and the electron concentration reads $\mu_d = 1.7\times
10^3$ cm$^2$/V\,s.
\begin{figure}[!h]
\includegraphics[width=0.95\linewidth]{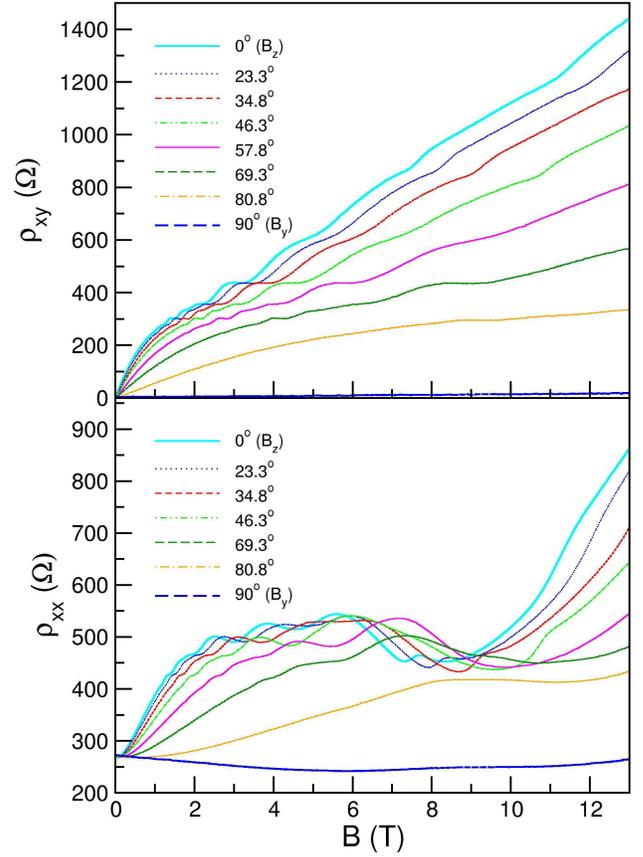}
\caption{\label{fig_3} 
(Color online) The experimental curves of the transversal 
$\rho_{xy}$ and longitudinal $\rho_{xx}$ magnetoresistance measured 
at $T = 0.3$~K for the set of tilt angles $\varphi$ of the applied 
magnetic field on the illuminated SL sample.}
\end{figure}
\begin{figure}[t]
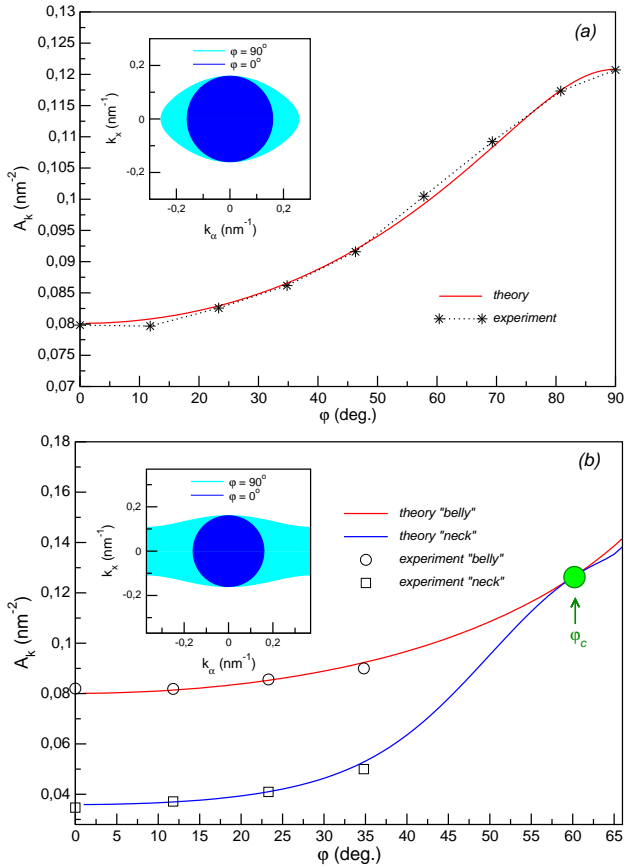

\begin{center}
\begin{tabular}{c}
\hbox{
\includegraphics[width=0.95\linewidth]{fig4.eps}} \\
\hbox{
\includegraphics[width=0.95\linewidth]{fig5.eps}}
\end{tabular}
\caption{\label{fig_4&5} 
(Color online) The experimental and theoretical extremal
cross-sections of the SL Fermi surface before {\it (a)} and after {\it(b)}
illumination against the tilt angle $\varphi$ between the sample and
the applied magnetic field. The green point on the lower figure
emphasizes the position of the critical angle $\varphi_c$ 
at which the ``belly'' and ``neck'' extremal cross-section areas 
become equal. Each inset shows two limiting cross-sections of the 
FS corresponding to the perpendicular ($\varphi = 0^{\circ}$) 
and parallel ($\varphi = 90^{\circ}$) orientations of the magnetic 
field with respect to the sample plane. The $k_{\alpha}$ on the 
horizontal axis denotes either $k_y$ 
(for $B = B_z$, $\varphi = 0^{\circ}$) or $k_z$ 
(for $B = B_y$, $\varphi = 90^{\circ}$).}
\end{center}
\end{figure}

Illumination changes dramatically both $\rho_{xx}$ and $\rho_{xy}$
magnetoresistance curves, as compared to those measured ``in
dark'' (see Fig.~\ref{fig_3}). 
Both curves became strongly nonlinear in the low-field region,
indicating that except of the electrons corresponding to the  ``belly''
orbits, new group of  electrons close to the ``neck'' orbits appeared. The
zero-field resistance $\rho_{xx}$ drops to less than one third of the
original value.  The SdH oscillations acquire more complicated form,
exhibiting two distinct periods evidently originated from the
``belly'' and ``neck'' orbits. The oscillations become weaker with
increasing the tilt angle $\varphi$ and disappear completely in strictly
in-plane magnetic fields, as it should be for the corrugated cylinder.
Hall plateaux appear in  $\rho_{xy}$ curves corresponding
to minima of the SdH oscillations in $\rho_{xx}$.
 
Due to the nonlinear shape of $\rho_{xy}$ in low magnetic fields, the
straightforward determination of the electron concentration $N_{il,H}$
becomes unreliable.  From the smooth part of $\rho_{xy}$ in higher
perpendicular magnetic fields (above $3$ T) we got $N_{il,H}\backsimeq
2.2\times 10^{11}$~cm$^{-2}$.

There are two angular dependences of extremal cross-sections,
corresponding to ``belly'' and ``neck'' orbits, determined from two
experimental periods of SdH oscillations. We were able to detect
oscillation periods only up to $\varphi\approx
35^{\circ}$. Cross-section areas of the ``belly'' and ``neck'' orbits
grow slowly for $\varphi<\varphi_c\,$, where 
$\varphi_c\approx 60^{\circ}$ is the critical angle at which 
both cross-sections become equal $\left(A_k^B=A_k^N\right)$.  From the
experimental angular dependences the parameters of the FS in
Eq.~(\ref{Ak_tilt}) can be determined by nonlinear curve fitting. We
have found $t=2$~meV, $E_F=10.5$~meV, which means that the SL miniband
width is $8$~meV and that the Fermi energy is $6.5$~meV above the top
of the miniband.

It follows from this value of $E_F$ and Eq.~(\ref{N_open}) that
$N_{il,SdH}\backsimeq 2.94\times 10^{11}$~cm$^{-2}$.  We consider this
number more reliable than $N_{il,H}\backsimeq 2.2\times
10^{11}$~cm$^{-2}$, determined from the high-field $\rho_{xy}$ data,
which obviously underestimates the true value. 

The zero-field resistance and $N_{il,SdH}$ give the mobility $\mu_{il}
= 3.6 \times 10^3$ cm$^2$/V\,s, higher than the mobility $\mu_d =
1.7\times 10^3$ cm$^2$/V\,s of the sample before illumination.

Good agreement between the experimental data, which stems from the
periods of SdH oscillations, and the theoretical curves based on the
cosine approximation of the SL miniband is illustrated 
in Fig.~\ref{fig_4&5}. It  confirms that this approach is
relevant for description of the electronic structure of the studied
sample.

\begin{figure}[t]
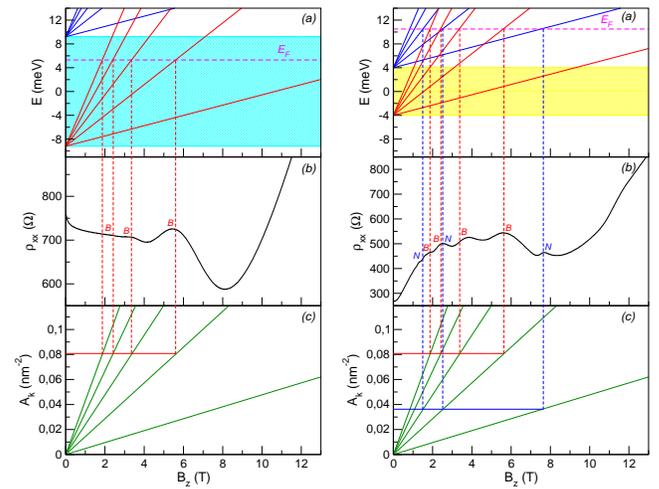

\begin{center}
\begin{tabular}{cc}
\hbox{
\includegraphics[width=0.475\linewidth]{fig6.eps}} &
\hbox{
\includegraphics[width=0.475\linewidth]{fig7.eps}}
\end{tabular}
\caption{\label{fig_6&7} 
(Color online){\it (a)} Electron energy structure of the SL ``in
dark'' (left figure) and illuminated one (right figure). The Landau
levels are plotted against the magnetic field applied perpendicular to
the sample. The blue fans  are associated with ``neck'' orbits
(and the top of the SL miniband), red ones with ``belly'' orbits (and the
bottom of the SL miniband). The tinged cyan and yellow regions show the SL
miniband ranges before and after illumination, respectively.  The dashed
lines denote the positions of the Fermi energy.  {\it (b)} The
longitudinal magnetoresistance curve $\rho_{xx}$ measured on the SL
before (left figure) and after illumination (right figure) in the
perpendicular magnetic field.  Letters {\it B} and {\it N} are
inscribed above the magnetoresistance oscillation maxima corresponding
either to a ``belly'' or ``neck'' orbit. {\it (c)} The cross-section
area of the SL Fermi surface as a function of the perpendicular magnetic field,
calculated on the basis of the Onsager-Lifshitz quantization rule
given by Eq.~(\ref{O-Lq}). }
\end{center}
\end{figure}

\begin{figure}[htb]
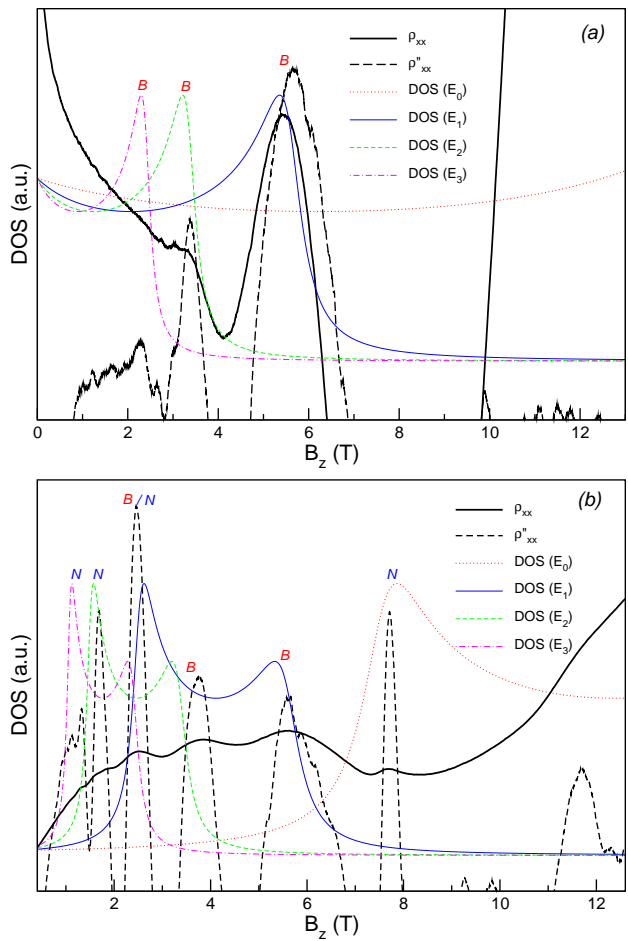

\begin{center}
\begin{tabular}{c}
\hbox{
\includegraphics[width=0.95\linewidth]{fig8.eps}} \\
\hbox{
\includegraphics[width=0.95\linewidth]{fig9.eps}}
\end{tabular}
\caption{\label{fig_8&9} 
(Color online) The longitudinal magnetoresistance $\rho_{xx}$ measured on the
SL ``in dark'' {\it (a)} and illuminated one {\it (b)} 
with their second derivatives compared to  
DOSs calculated as functions of the magnetic field applied
perpendicular to the sample.
The DOSs calculated only for the lowest four eigenenergies are shown.}
\end{center}
\end{figure}

Good consistency of our simple theory with experiments is further
confirmed by data obtained in strictly perpendicular magnetic fields,
which are presented in Fig.~\ref{fig_6&7}. Besides the longitudinal
magnetoresistances $\rho_{xx}$, the fans of Landau levels corresponding to the
``belly'' and ``neck'' orbits are shown, together with the calculated
minibands and Fermi energies of both as-grown and illuminated
samples. Intersects of Landau levels with calculated $E_F$ coincide with
positions of oscillation maxima and thus give the values of
cross-section areas which are in excellent agreement with experimental
values. The graphical method of comparison of the theory and experiment
is very useful as it allows to distinguish whether a peak of the
longitudinal magnetoresistance belongs to the ``belly'' or the
``neck'' extremal orbit.

Experimentally determined parameters, $t$ and $E_F$, have also been
used to calculate the density of states (DOS) for several lowest
Landau subbands of the SL before and after illumination.  The
resulting DOSs are shown in Fig.~\ref{fig_8&9} together with
corresponding longitudinal magnetoresistance curves and their second
derivatives as functions of the perpendicular magnetic field. The
reasonable agreement of the oscillation peak positions with the DOS
extrema
supports our interpretation.
\section{Discussion \label{disc}}
We have measured low-temperature longitudinal and transversal
magnetoresistance of the short-period SL with Si-doped
barriers in the as-grown state and after illumination.

In the as-grown state, the SL behaves as a 3D system, exhibiting a
single period SdH oscillations for all orientations of the external
magnetic field. This implies that the neighboring QWs are
strongly coupled and electrons forming an anisotropic 3D gas are
present with nonzero probability not only in the QWs but
also inside the barriers. It is possible only if all shallow Si donors
are ionized and empty shallow levels lie in the thermodynamic
equilibrium well above the Fermi energy. Thus the depleted layer is
extended through the whole Al$_{0.3}$Ga$_{0.7}$As barrier and the
Fermi energy is not fixed in Al$_{0.3}$Ga$_{0.7}$As by the donor
level as in the standard modulation-doped structures.

The situation becomes more complicated after the sample illumination,
which leads to the persistent photoconductivity and increases
persistently the density of carriers. The origin of the PPC is 
attributed to ionization of deep Si donor states in the 
Al$_{0.3}$Ga$_{0.7}$As barrier. As the original ``dark'' value of the
concentration and the magnetoresistance can be recovered by heating
the sample to temperatures above 100~K, the illuminated sample is in a
reversible metastable state, separated from the thermodynamic
equilibrium by the capture barrier which prevents the return of
electrons back to ionized Si donor states at low temperatures.  

In spite of year-long extensive studies the nature of deep levels and the
capture barrier is still a subject of discussions.
The detailed discussion of the properties of deep donor levels is
beyond the scope of this publication, the main alternative
explanations of the PPC and the photoionization process can be found, e.g., in 
Refs.~\onlinecite{chadi,feng,jantsch,shayegan}.

Here we  consider as important that the capture barrier is formed around the
ionized donors inside the doped Al$_{0.3}$Ga$_{0.7}$As layer. 
It can be a combined effect of both the band-offset between
Al$_{0.3}$Ga$_{0.7}$As and GaAs and this capture barrier which
leads to suppression of the inter-well coupling and consequently to
the reduction of the miniband width.
\section{Conclusions \label{Concl}}
In conclusion, longitudinal and transversal magnetoresistances have
been measured in tilted magnetic fields on the
GaAs/AlGaAs SL sample. Illumination of the
sample enhances the electron concentration in the QWs and makes the
SdH oscillations more complicated. The results have been analyzed in
terms of a simple tight-binding model. The effect of the illumination
can be described as a 3D$\rightarrow$2D transition, where the FS of
the system changes from closed ovals contained within the first
Brillouin zone to an open corrugated cylinder, characteristic for a
system composed of weakly coupled 2D electron layers.  The
illumination thus not only enhances the electron concentration in the
layers, but simultaneously suppresses the strength of the inter-layer
coupling as indicated by the decrease of the SL miniband width. 
This rather surprising result can be qualitatively explained
in terms of increasing the effective barrier height due to a
photon-induced ionization of deep donor impurity levels within
Al$_{0.3}$Ga$_{0.7}$As barriers. 


\section{Acknowledgements}
This work has been supported by the European Commission contract
No. RITA-CT-2003-505474, Ministry of Education of the Czech Republic
Center for Fundamental Research LC510 and Academy of Sciences of the
Czech Republic project KAN400100652.

\end{document}